\documentclass[twocolumn,amsmath,amssymb,prl,showpacs,showkeys,superscriptaddress]{revtex4}
\usepackage{amssymb}
\usepackage[]{graphics}
\usepackage{epsf}
\usepackage{calc}
\usepackage{epsfig}
\usepackage[]{color}
\usepackage[pagebackref=true]{hyperref}
\usepackage[collision]{chemsym}
\begin{document}

\title{Fabrication of high quality ferromagnetic Josephson junctions}

\author{M. Weides}
 \affiliation{CNI - Center of Nanoelectronic Systems for Information
Technology, Research Centre Juelich, D-52425 Juelich, Germany}
 \email{m.weides@fz-juelich.de}
\author{K. Tillmann}
\affiliation{Institute for Solid State Research, Research Centre
Juelich, D-52425 Juelich, Germany} \affiliation{Ernst Ruska-Centre
for Microscopy and Spectroscopy with Electrons, Research Centre
Juelich, D-52425 Juelich, Germany}
\author{H. Kohlstedt}
 \affiliation{CNI - Center of Nanoelectronic Systems for Information
Technology, Research Centre Juelich, D-52425 Juelich, Germany}
\affiliation{Department of Material Science and Engineering and
Department of Physics, University of Berkeley, California 94720,
USA}

\date{\today}


\begin{abstract}
 We present ferromagnetic  $\Nb/\Al_2\O_3/\Ni_{60}\Cu_{40}/\Nb$
Josephson junctions (SIFS) with an ultrathin $\Al_2\O_3$ tunnel
barrier. The junction fabrication was optimized regarding junction
insulation and homogeneity of current transport. Using
ion-beam-etching and anodic oxidation we defined and insulated the
junction mesas. The additional 2 nm thin Cu layer below the
ferromagnetic $\Ni\Cu$ (SINFS) lowered interface roughness and
ensured very homogeneous current transport. A high yield of
junctional devices with $j_c$ spreads less than 2\% was obtained.
\end{abstract}

 \maketitle

\section{Introduction}
The realization of qubits for quantum computation attracts
considerable interest. One approach is to use low $T_c$ Niobium
Josephson junctions (JJ) which utilize an ultrathin ferromagnetic
layer to change the coupling of the two superconducting electrodes
\cite{buzdinJETP,Demler}. Whether the junction is in the $0$ or
$\pi$ coupled state depends on the properties of ferromagnetic layer
and especially its thickness $d_F$. The coupling state can be
determined by
\begin{displaymath} \label{IcRn} I_c \propto \exp(-d_F/\xi_{F1}) |\cos\frac{d_F}{\xi_{F2}}+\frac{\xi_{F1}}{\xi_{F2}}\sin \frac{d_F}{\xi_{F2}}|\end{displaymath}
with characteristic decay $\xi_{F1}$ and oscillation length
$\xi_{F2}$ \cite{Buzdin2005}. Transition from $0$ to $\pi$ occurs at
$Ic(d_{F,0-\pi})=0$ when critical current $\I_c$ cancels out. Such
JJs which can be realized either in an SFS
\cite{RyazanovPi_Coupling,sellier_T_induced_0_pi_state,blumNb_Cu_Ni_Nb,surgers_Strunk_nanostructured_SFS}
or an SIFS \cite{kontos_negative_coupling} layer sequence (S:
Superconductor, F: Ferromagnet, I: Insulator). In contrast to SFS
junctions the area-resistance ($R\times A$) product of SIFS
structures can be tuned over orders of magnitude by using
appropriate oxidation conditions for the tunnel barrier formation.
High ($R\times A$) products simplify the voltage readout.\\ The
first step to realize a qubit is the formation of a semifluxon.
Recent theoretical considerations predict semifluxons in SIFS
junctions with a step-like ferromagnetic layer
\cite{goldobin_Semifluxons_LJJ}. Exploiting such semifluxon for
qubits systems \cite{Goldobin_quantum_tunneling} demands low spread
of extrinsic and intrinsic parameters, e.g. junction area and
current density.
\section{Samples}
\begin{figure}[b]
\begin{center}
\includegraphics[width=\textwidth/2-1cm]{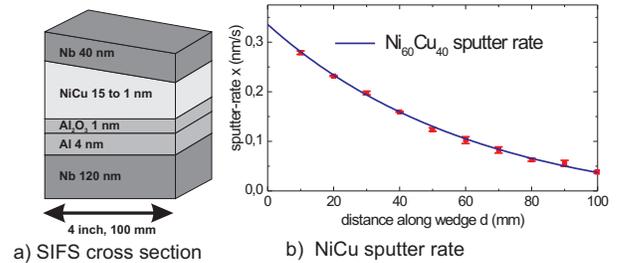}
\caption{\label{SIFS_wedge} (a) SIFS stack with "wedge"-shaped
F-layer and (b) decreasing of the $\Ni\Cu$-sputter rates across a
$4$ inch wafer}
\end{center}
\end{figure}

The fabrication of SIFS multi-layers was performed in-situ by a
computer-controlled Leybold Univex 450B magnetron sputtering system.
Thermally oxidized $4$-inch Si wafer served as substrates. The
wafers were clamped onto a water cooled $\Cu$-block. First a $120$
nm $\Nb$ bottom electrode and an $5$ nm thick $\Al$ overlayer were
deposited. Subsequently the aluminium was oxidized for $30$ min in a
separate chamber at an oxygen partial pressure ranging from $0.015$
to $0.45$ mbar. As ferromagnetic layer we deposited the diluted
$\Ni_{60}\Cu_{40}$ alloy ($T_C=225\:\rm{K}$), followed by a $40$ nm
$\Nb$ cap-layer. For realization of "wedge" shaped $\Ni\Cu$-layer
(Fig. \ref{SIFS_wedge}), the substrate and sputter target are
shifted about half the substrate length. This facilitates the
preparation of SIFS (and SINFS, N: normal metal) junctions with
different F-layer thicknesses to avoid the inevitable run to run
variations. In this way we prepared F-layers ranging between $1$ and
$15$ nm.

The argon pressure during sputtering was $7.0 \cdot 10^{-3}\:
\rm{mbar}$ for $\Nb \: {\rm and} \: \Al$ and $4.2 \cdot 10^{-3}
\rm{mbar}$ for $\Cu \: {\rm and} \: \Ni\Cu$, respectively. The
background pressure was $5 \cdot 10^{-7}\: \rm{mbar}$. Niobium was
deposited statically with $2.00 \:\rm{nm/s}$ at a power density of
$5\; {\rm W/cm^2}$ and $\Ni\Cu$: $0.34 \:\rm{nm/s}$ at $0.6 \: {\rm
W/cm^2}$, while $\Al$ and $\Cu$ were deposited during sample
rotation and at much lower deposition rates to obtain very
homogeneous and uniform films: $\Al$: $0.05 \:\rm{nm/s}$ and $\Cu$:
$0.1 \:\rm{nm/s}$, both at $1.9\: \rm{W/cm^2}$.

\begin{figure}
\begin{center}
\includegraphics[width=\textwidth/2-1cm]{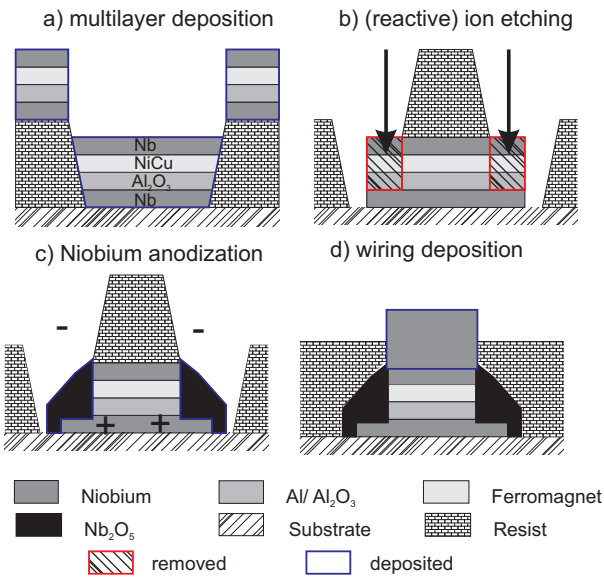} \caption{\label{patterningSINFS} (a) to (d) Three level
photo mask procedure including ion-etching and Nb-anodization}
\end{center}
\end{figure}

\subsection{Patterning}
\begin{figure}[t]
\begin{center}
\includegraphics[width=\textwidth/2-1cm]{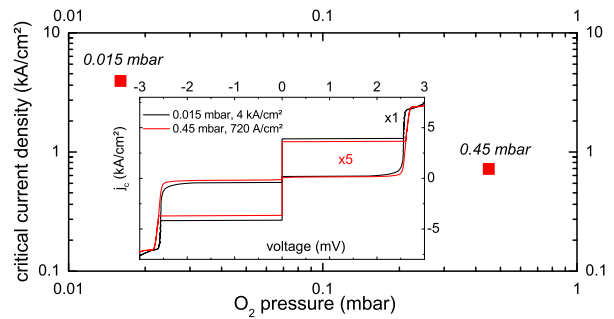}
\caption{\label{UebersichtOxidationp_Vs_jc} SIS junctions with
different oxidation conditions for the tunnel barrier forming.}
\end{center}
\end{figure}

Tunnel junctions with a crossbar geometry were patterned using
optical lithography and $\Ar$ ion beam milling. A three level photo
mask procedure was applied. First the bottom SIFS-layer areas were
defined by a lift-off process. The situation after deposition of the
SIFS sequence is shown in Fig. \ref{patterningSINFS} (a).\\After the
lift-off various kinds of tunnel junctions were defined by applying
the second photo mask step followed by reactive ion beam etching for
the $\Nb$ layer and $\Ar$ for the $\Ni\Cu$ and $\Al$ layers. The
etching was controlled by a mass spectrometer and the procedure was
stopped by reaching the $\Al_2\O_3$ tunnel barrier (Fig.
\ref{patterningSINFS}(b)). During etching the substrate was tilted
by $70^\circ$ and rotated to avoid etch fences at the edges of the
mesas. The mesas were isolated by SNEAP (Selective Niobium
Anodization Process) \cite{Gurvitch_SNEAP}, Fig.
\ref{patterningSINFS}(c).\\ It is interesting to note that the
anodization was successful in the presence of $\Ni\Cu$ layer. We
obtained no problems with parallel currents through the $\Ni\Cu$
layer during anodization. Probably the ferromagnetic layer is so
thin that it is immediately overgrown by $\Nb_2\O_5$ and $\Al_2\O_3$
shortly after starting the anodization procedure. At a rate about
$1\: \rm{V/s}$ we anodized the junction up to a voltage of $60 \:
\rm{V}$ (corresponding to $51 \: \rm{nm}$ of anodized Niobium) was
reached. The form factor of $2.3$ for Niobium oxidation corresponds
to $120 \: \rm{nm}$ of formed $\Nb_2\O_5$, providing a complete side
coverage of the barrier and the ferromagnetic layer.\\ In the last
photo mask step the wiring layer was defined. After a slight ion
beam etching to achieve low contact resistance, a $300 \: \rm{nm}$
thick $\Nb$ wiring was deposited. In Fig. \ref{patterningSINFS} (d)
the schematic
cross-section of the device is shown.\\
To check the procedure standard SIS junction were fabricated with
areas between $25 \: \rm{\mu m^2}$ and $1000 \: \rm{\mu m^2}$. In
Fig. \ref{UebersichtOxidationp_Vs_jc} two SIS characteristics are
shown. For the ferromagnetic SIFS and SINFS junctions the areas were
$10.000 \: \rm{\mu m^2}$ large.

\subsection{Oxidation}
Besides a reliable junction patterning the interlayer roughness is
essential for high quality ferromagnetic junctions. For SIFS
junctions we optimized the $\Al_2\O_3$ roughness as well as the
roughness of the $\Ni\Cu$ alloy to achieve low spreads of $j_c$ from
run to run. In this work we kept the oxidation time constant ($30$
min) and varied the oxygen pressure between $0.015$ and $0.45$ mbar.
Using the very low oxidation pressure ($0.015$ mbar) we obtained
current densities $j_c$ of $4\: \rm{kA/cm^2}$ (Fig.
\ref{UebersichtOxidationp_Vs_jc}) to counterbalance the strong
Cooper-pair breaking in the ferromagnetic alloy, as seen in the
$I_c(d_F)$ dependence \cite{SIFS_paper}.

\subsection{SIFS and SINFS}
\begin{figure}[b]
\begin{center}
\includegraphics[width=\textwidth/2-1cm]{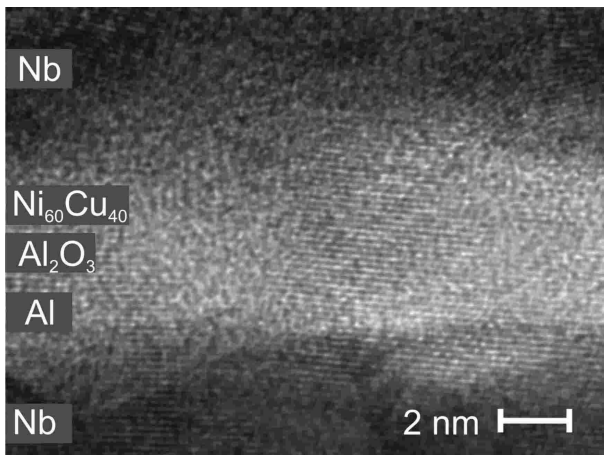}
\caption{\label{TEM_SIFS}Cross-sectional HRTEM image of a
$\Nb/\Al/\Ni\Cu/\Nb$ stack}
\end{center}
\end{figure}

On reference samples we performed ex-situ atomic force microscopy
(AFM) measurements. The roughness of the $\Si\O_2$ surface was less
than $0.3 \:\rm{nm}$ (rms). A $120\:\rm{nm}$ $\Nb$ film showed a
roughness of $0.44 \:\rm{nm}$, which increased to $0.6 \:\rm{nm}$
after deposition of $5 \:\rm{nm}$ $\Al$ and even up to
$0.9\:\rm{nm}$ after further $4\:\rm{nm}$ of $\Ni\Cu$. Insertion of
a $2\:\rm{nm}$ $\Cu$ film after oxidation decreases the roughness of
the SIN stack down to $0.50\:\rm{nm}$. Now the top roughness of SINF
(F: $4.7\:\rm{nm}$) stack is about $0.68\:\rm{nm}$. The
ferromagnetic interlayer in a SINFS stack exhibited at both
interfaces of $\Ni\Cu$ a lower roughness than in a SIFS stack.
Non-uniform growth of $\Ni\Cu$ on top $\Al_2\O_3$ may be causes by
island-formation of the first
monolayers when grown directly on $\Al_2\O_3$.\\
Fig. \ref{TEM_SIFS} displays a cross-sectional HRTEM
(High-Resolution Transmission Electron Microscopy) image of a SIFS
stack which was taken under bright atom contrast conditions using a
spherical-aberration corrected electron microscope \cite{Tillmann}.
Because of the polycrystalline structure of the layers, individual
nanocrystallites are not uniformly oriented with respect to the
incident electron beam effecting that only certain lattice planes
are resolved at atomic plane distances. Moreover, the image contrast
is slightly bleary, which is most presumably due to both, the
formation of amorphous interlayers during deposition and electron
beam damage during operation of the instrument.

However, the basic layer structure becomes clearly visible from this
image, i.e. the $\Al$ layer appears undulated, as opposed to a
potential two-dimensionally flat layer. The lower $\Nb-\Al$ and the
upper $\Ni\Cu-\Nb$ interfaces show the very same undulations,
meaning that roughness and interdiffusion occurs on a small length
scale. Nonetheless the $\Al$, $\Al_2\O_3$ and $\Ni\Cu$ layers are
not directly distinguishable from each other in the micrograph. We
attribute this observation to similar electron scattering amplitudes
of the former materials and the interlayer roughness at rather small
layer thicknesses, which may give rise to projection artefacts.

\section{Results and discussion}
Our approach to patter ferromagnetic JJs differed from the standard
procedure
\cite{RyazanovPi_Coupling,sellier_T_induced_0_pi_state,blumNb_Cu_Ni_Nb,surgers_Strunk_nanostructured_SFS,kontos_negative_coupling}
where junction insulation is done by deposition of silicon-oxides
after etching. The $\Nb_2\O_5$ exhibits nearly defect free
insulation between superconducting electrodes, even for thick (15
nm) ferromagnetic interlayers.

\begin{figure}[b]
\begin{center}
\includegraphics[width=\textwidth/2-1cm]{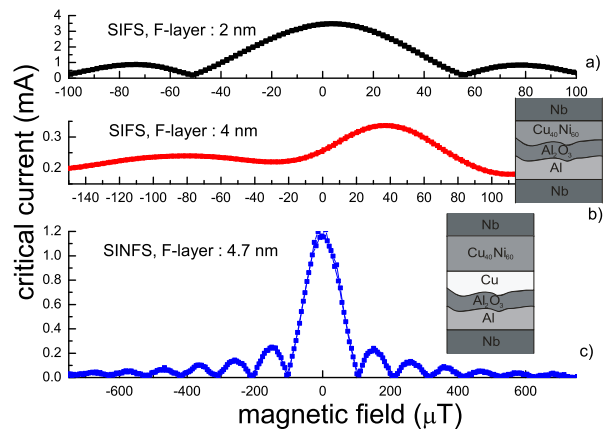} \caption{\label{IcH_SIFS} $I_c(H)$
of (a) SIFS (2 nm $\Ni\Cu$), (b) SIFS (4 nm $\Ni\Cu$) and (c) SINFS
($\Cu$ 2 nm, $\Ni\Cu$ 4.7 nm) stack. Oxygen pressure is 0.45 mbar
for SIFS and 0.015 mbar for SINFS type.}
\end{center}
\end{figure}

Electric transport measurements were made in a liquid Helium dewar
at $4.2$ K. In Fig. \ref{UebersichtOxidationp_Vs_jc} the current
voltage characteristics for SIS junctions at different oxidation
pressures are shown. The figures of merit $V_m=I_cR(2\rm{mV})$,
McCumber-Stewart parameter $\beta_c =\frac{2eI_cR^2_nC}{\hbar}$ and
$\V_c=I_cR_n$ are for the $j_c=4\:\rm{kA/cm}^2$ junction:
$V_m=12\:\rm{mV},\: \beta_c=3.10, V_c=1.60\:\rm{mV}$ and for the
$j_c=720\:\rm{kA/cm}^2$ junctions: $V_m=32.1 \:\rm{mV},
\:\beta_c=14.8, V_c=1.49\:\rm{mV}$. Larger $j_c$ corresponded to
inhomogeneous tunnel barrier formation, hence increased subgap
leakage current and a decreased energy gap (inset of figure
\ref{UebersichtOxidationp_Vs_jc}). Still these higher $j_c$
junctions exhibit good junction parameters, so we used their tunnel
barrier oxidation conditions ($0.015 \: \rm{mbar}$) for the
fabrication of SINFS stacks.

The Fraunhofer modulation of $I_c$ is seen in Fig. \ref{IcH_SIFS}
for SIFS junctions with $2 \:\rm{nm}$ (a) and $4\:\rm{nm}$ (b)
$\Ni\Cu$ layers and a SINFS junction with $2\:\rm{nm}$ $\Cu$ and
$4.7\:\rm{nm}$ $\Ni\Cu$(c). All the ferromagnetic junction
investigated in this work are still in $0$ coupled regime
\cite{SIFS_paper}. In general the interface barrier roughness leads
to inhomogeneous current transport, which can cancel out the
coherent Josephson coupling and leads to disturbed Fraunhofer
pattern. The conditions for non-uniform supercurrent are given by
the ratio of $\Ni\Cu$ interface roughness over the decay length
$\xi_{F1}$ and oscillation length $\xi_{F2}$ of the supercurrent. In
$\Ni_{60}\Cu_{40}$ these values are about $1-3\: \rm{nm}$, to be
published in \cite{SIFS_paper}.\\For thin (below $3\:\rm{nm}$)
$\Ni\Cu$ layers we see a clean Fraunhofer modulation (Fig.
\ref{IcH_SIFS} (a)). However for thicker $\Ni\Cu$ layers the
$I_c(H)$ deviated considerably (Fig. \ref{IcH_SIFS} (b)). Although
electrical measurements on SIS junctions suggested a high quality
barrier (inset of Fig. \ref{UebersichtOxidationp_Vs_jc}), this
suggests some finite roughness of the $\Al_2\O_3$ tunnel barrier
and/or of the $\Ni\Cu$ layer. As AFM-measurements indicated the
$2\:\rm{nm}$ $\Ni\Cu$ layer on-top the $\Al_2O_3$ forms similar
roughness contours, so effective thickness of $\Ni\Cu$ is constant
and super-current transport remains homogeneous as seen in Fig.
\ref{IcH_SIFS} (a). When doubling the thickness of $\Ni\Cu$ to
$4.7\:\rm{nm}$ its top roughness is about $0.9\: \rm{nm}$ and the
effective $\Ni\Cu$ layer thickness is not uniform. This is supported
by transport measurements seen in Fig. \ref{IcH_SIFS} (b). By
including a $2\: \rm{nm}$ $\Cu$ layer under the $\Ni\Cu$ the
$I_c(H)$ we recovered the clean Fraunhofern pattern. We could even
use the thinner $\Al_2\O_3$ barrier for SINFS, although this should
increase the $\Al_2\O_3$ roughness slightly. The $\Cu$-layer between
$\Al_2\O_3$ and $\Ni\Cu$ smoothes the lower $\Ni\Cu$-interface and
provides an uniform effective ferromagnetic layer. Even for thicker
(up to $8.5\:\rm{nm}$) $\Ni\Cu$-layers we see uniform supercurrents
through such SINFS junctions (fig. \ref{IcH_SIFS} (c)). For $\Ni\Cu$
thicker than $8.5\:\rm{nm}$ we can not measure any supercurrent due
to the strong Cooper pair breaking.
This will be reported elsewhere \cite{SIFS_paper}.\\
\begin{figure}[]
\begin{center}
\includegraphics[width=\textwidth/2-1cm]{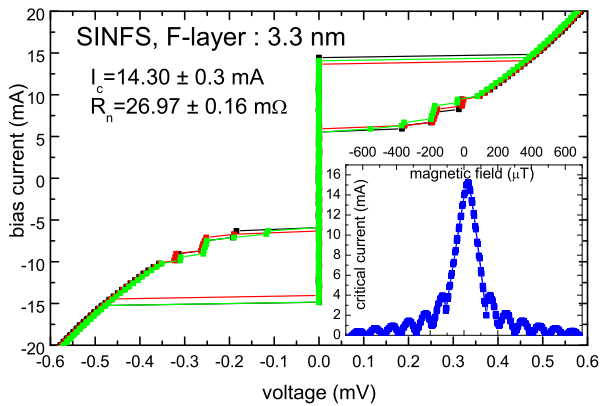} \caption{\label{SFS31_J}IVs of
three SINFS junction, $\O_2$ oxidation pressure is 0.015 mbar.
$\delta I_c=2\%$, $\delta R_n= 0.5\%$}
\end{center}
\end{figure}

The strong proximity effect of $\Cu$ leads to weak pair breaking of
the super-current in the N-layer. SIFS and SINFS samples (N: $2$ nm
Cu layer) showed identical critical current densities, so junction
properties are determined by the ferromagnetic layer. For all
$\Ni\Cu$ thicknesses we obtained a low junction to junction
deviation. In Fig. \ref{SFS31_J} the characteristics of three
under-damped SINFS junctions with a $3.3\: \rm{nm}$ $\Ni\Cu$ layer
are shown. The parameter spread of critical current and normal
resistance (and therefore $\beta_c$) is below $2\rm{\%}$
($I_c=14,3\: \rm{mA}$, $R_n=26.97\:\rm{m\Omega}$ and $\beta_c=5.0$).
Even the very sensitive sub-gap characteristics are nearly identical
for all junctions as seen in Fig. \ref{SFS31_J}. These junction
exhibit a high-quality $I_c(H)$ pattern (inset of Fig.
\ref{SFS31_J}), just like the Fraunhofern pattern of SINFS with $4.7
\:\rm{nm}$ $\Ni\Cu$ layer in Fig. \ref{IcH_SIFS} (c).

\section{Conclusion}
Motivated by the demand for ferromagnetic JJ with low parameter
spread, we have developed an alternative fabrication process.
Transport measurements on SINFS junction showed that the quality of
junctions was considerable improved using $\Nb_2\O_5$ as insulator
and planarization of the ferromagnetic interlayer by an additional
$\Cu$ layer. Our fabrication procedure may offer a solution for the
strict uniformity requirements for the formation of
qubits.\textcolor[rgb]{1.00,1.00,0.00}{}

\bibliographystyle{elsart-num}

\bibliography{references}



\end{document}